\documentclass{osa-article}
\journal{oe}
\usepackage[utf8]{inputenc}
\newcommand{\opt}{\text{opt}}
\newcommand{\SPD}{\text{SPD}}

\usepackage{amsmath}
\usepackage{array}
\usepackage{graphicx}
\usepackage{xcolor}
\begin{document}
\title{Spatially multiplexed single-photon sources based on incomplete binary-tree multiplexers}
\author{Peter Adam,\authormark{1,2,*} Ferenc Bodog\authormark{3}, and Matyas Mechler\authormark{2}}

\address{\authormark{1}Institute for Solid State Physics and Optics, Wigner Research Centre for Physics,\\ P.O. Box 49, H-1525 Budapest, Hungary\\
\authormark{2}Institute of Physics, University of P\'ecs, Ifj\'us\'ag \'utja 6, H-7624 P\'ecs, Hungary\\
\authormark{3}MTA-PTE High-Field Terahertz Research Group, H-7624 P\'ecs, Hungary}

\email{\authormark{*}adam.peter@wigner.hu}

\begin{abstract}
We propose two novel types of spatially multiplexed single-photon sources based on incomplete binary-tree multiplexers.
The incomplete multiplexers are extensions of complete binary-tree multiplexers, and they contain incomplete branches either at the input or at the output of them.
We analyze and optimize these systems realized with general asymmetric routers and photon-number-resolving detectors by applying a general statistical theory introduced previously that includes all relevant loss mechanisms.
We show that the use of any of the two proposed multiplexing system can lead to higher single-photon probabilities than that achieved with complete binary-tree multiplexers.
Single-photon sources based on output-extended incomplete binary-tree multiplexers outperform those based on input-extended ones in the considered parameter ranges, and they can in principle yield single-photon probabilities higher than 0.93 when they are realized by state-of-the-art bulk optical elements.
\end{abstract}

\section{Introduction}
The development of single-photon sources (SPSs) is of utmost importance for the effective realization of a number of experiments in the fields of photonic quantum technology and quantum information processing~\cite{EisamanRSI2011, MScott2020}.
A promising realization of periodic SPSs are the heralded single-photon sources that can yield highly indistinguishable single photons in near-perfect spatial modes with known polarization \cite{PittmanOC2005, Mosley2008, Brida2011, RamelowOE2013, MassaroNJP2019}.
In such sources, the detection of one member of a correlated photon pair generated in spontaneous parametric down-conversion (SPDC) or spontaneous four-wave mixing (SFWM) heralds the presence of its twin photon.
The inherent probabilistic nature of the photon pair generation in these nonlinear sources results in the occasional occurrence of multiphoton events in the pair generation. This detrimental effect can be reduced by various multiplexing techniques such as spatial multiplexing \cite{Migdall2002, ShapiroWong2007, Ma2010, Collins2013, Meany2014, Francis2016, KiyoharaOE2016} and time multiplexing \cite{Pittman2002, Jeffrey2004, Mower2011, Schmiegelow2014, Francis2015, Kaneda2015, Rohde2015, XiongNC2016, Hoggarth2017, HeuckNJP2018, Kaneda2019, Lee2019, MagnoniQIP2019} where heralded photons generated in a set of multiplexed units realized in space or  in time are rerouted to a single output mode by a switching system.
In multiplexed SPSs, the multi-photon noise can be suppressed by keeping the mean photon number of the generated photon pairs low in a multiplexed unit, while the high probability of successful heralding in the whole system can be guaranteed by the use of several multiplexed units.
Multi-photon events can also be reduced by using single-photon detectors with photon-number-resolving capabilities for heralding \cite{RamelowOE2013, BonneauNJP2015, KiyoharaOE2016, Bodog2020}.
High-efficiency inherent photon-number resolving detectors (PNRDs) in various realizations such as transition edge sensors \cite{Lita2008, Fukuda2011, Schmidt2018, Fukuda2019} or superconducting nanowire detectors~\cite{Divochiy2008, Jahanmirinejad2012} are also available for this task.

An unavoidable issue of real multiplexed systems is the appearance of various losses of the optical elements in both the heralding stage and the multiplexing system that leads to the limitation of the performance of multiplexed SPSs \cite{MazzarellaPRA2013, BonneauNJP2015}.
Full statistical frameworks have already been developed for the description of any kind of multiplexed SPSs using various photon detectors that take all relevant loss mechanisms into account \cite{Adam2014, Bodog2016, Bodog2020}.
These frameworks make it possible to optimize multiplexed SPSs, that is, to maximize the output single-photon probability by determining the optimal values of the system size and the mean number of photon pairs generated in the multiplexed units for a given set of loss parameters.
The analysis of various multiplexed SPSs showed that the single-photon probability that can be achieved in these systems after the optimization are different even by using identical optical elements in the setups.
This finding motivates the development of novel multiplexing schemes leading to higher single-photon probabilities.

In spatial multiplexing, which is the topic of the current research, several individual pulsed heralded SPSs are used in parallel.
In these systems, after a successful heralding event in one of the multiplexed units, the heralded signal photon is rerouted to the single output by a set of binary photon routers.
In the literature, these routers were proposed to be arranged into an asymmetric (chain) or a symmetric (binary-tree) structure \cite{MazzarellaPRA2013, BonneauNJP2015}.
Spatial multiplexing has been realized experimentally up to two multiplexed units by using SFWM in photonic crystal fibers \cite{Collins2013, Francis2016}, and up to four multiplexed units by using SPDC in bulk crystals \cite{Ma2010,KiyoharaOE2016} and waveguides \cite{Meany2014}. In all these experiments symmetric, that is, complete binary-tree multiplexers were used.

In this paper, we propose two novel types of spatially multiplexed SPSs based on incomplete binary-tree multiplexers built with asymmetric binary routers.
We analyze the proposed systems in detail by applying the statistical theory introduced in Ref.~\cite{Bodog2020} for describing multiplexed SPSs equipped with PNRDs.
We show that the proposed schemes can yield higher single-photon probabilities than the complete binary-tree scheme realized thus far in experiments.

\section{Incomplete binary-tree multiplexers}\label{sec:mplxers}
The idea of any spatial multiplexing scheme is to convey heralded photons generated in a set of multiplexed units (MUs) to the single output of a multiplexer characterized by a given geometry of a set of binary (2-to-1) photon routers (PRs).
Each of the multiplexed units contain a nonlinear photon pair source, a detector used to herald the presence of a signal photon of a photon pair by detecting the corresponding idler (twin) photon, and optionally a delay line placed in the path of the signal photon.
Using such delay lines might be required in order to introduce a sufficient delay into the arrival time of the signal photons to the inputs of the multiplexer arms, and thus to enable the operation of the logic controlling the routers.
When the detector is activated by the presence of one or more idler photons of photon pairs generated by the nonlinear photon pair source in a multiplexed unit, the corresponding signal photons are coupled into the multiplexer.
In the most general case, photon-number-resolving detectors (PNRDs) can be used to detect the idler photons.
These detectors can realize any detection strategy defined by the actual detected photon numbers, that is, the set of predefined number of idler photons for which the corresponding signal photons generated in the nonlinear process are allowed to enter the multiplexer.
As for the periodicity of the single-photon source required in most of the applications, it can be guaranteed by pulsed pumping of the photon pair source.

\begin{figure*}[bt]
    \centering
    \includegraphics[width=0.3\textwidth]{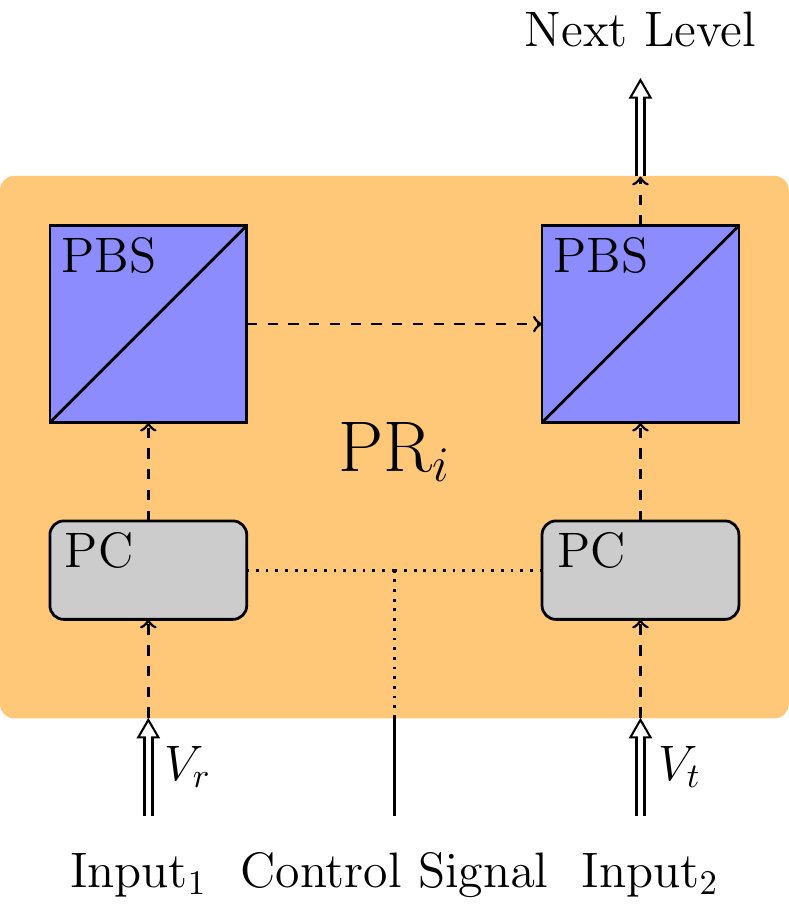}
\caption{Schematic figure of a bulk optical photon router PR$_i$. PBSs denote polarizing beam splitters, PCs are Pockels cells. $V_r$ and $V_t$ denote transmission coefficients characterizing the losses for Input$_1$ and Input$_2$, respectively.
\label{fig:Router}}
\end{figure*}
After the generation of photon pairs and the detection of the idler photons, the corresponding signal photons are conveyed into a multiplexer system characterized by a particular arrangement of a set of photon routers (PRs).
In our analysis, all the PRs forming the spatial multiplexer are assumed to be identical.
Routers are usually assumed to be symmetric, however, this restriction is not necessary, routers can be asymmetric, that is, they might have different transmission coefficients assigned to their two input ports.
Figure~\ref{fig:Router} presents a possible bulk optical realization of such an asymmetric binary photon router.
The building blocks of these routers are two Pockels-cells (PCs) serving as possible entrance points of the signal photons generated in the multiplexed units, and two polarizing beam splitters (PBSs), one of which acting as the output of the PR.

The polarization of the signal photons is known at the two input ports of the router. The PCs controlled by a priority logic can modify the polarization of these photons so that the PBSs can select and reroute the photons in the chosen mode to the output of the routers and eventually to the output of the whole multiplexer. If a mode is not selected, it can be directed out of the system or it can be absorbed by a suitable optical element.

Asymmetric binary routers are characterized by two transmission coefficients $V_r$ and $V_t$ corresponding to the transmission probabilities of the photons entering the router at Input$_1$ and Input$_2$, respectively. In the case of the router presented in Fig.~\ref{fig:Router} $V_r$ quantifies the losses due to the transmission through a PC and the two reflections in the PBSs. The other transmission, $V_t$, describes the losses introduced by the transmission through a PC and a PBS. These transmissions also contain an additional propagation loss in the router. Later on in this paper $V_t$ and $V_r$ will be referred to as the \emph{transmission and reflection efficiencies} and in all the schemes discussed in our work we will use routers with the coefficients $V_r$ and $V_t$ belonging to the left and right inputs of the router, respectively.

Previous papers aiming at theoretical modeling or experimental realization of periodic single-photon sources based on spatial multiplexing used two main types of multiplexers.
One of them is an asymmetric architecture in which the routers are arranged into a chain structure, that is, the outputs of the newly added routers are always coupled to one of the inputs of the previously added router \cite{MazzarellaPRA2013, BonneauNJP2015}.
The other geometry is a symmetric structure in which the constituent routers are arranged into a complete binary-tree multiplexer (CBTM) \cite{Migdall2002, ShapiroWong2007, Ma2010, Collins2013, Meany2014, BonneauNJP2015, Francis2016, KiyoharaOE2016, Adam2014, Bodog2016, Bodog2020}.

\begin{figure}
    \centering
    \includegraphics[width=\textwidth]{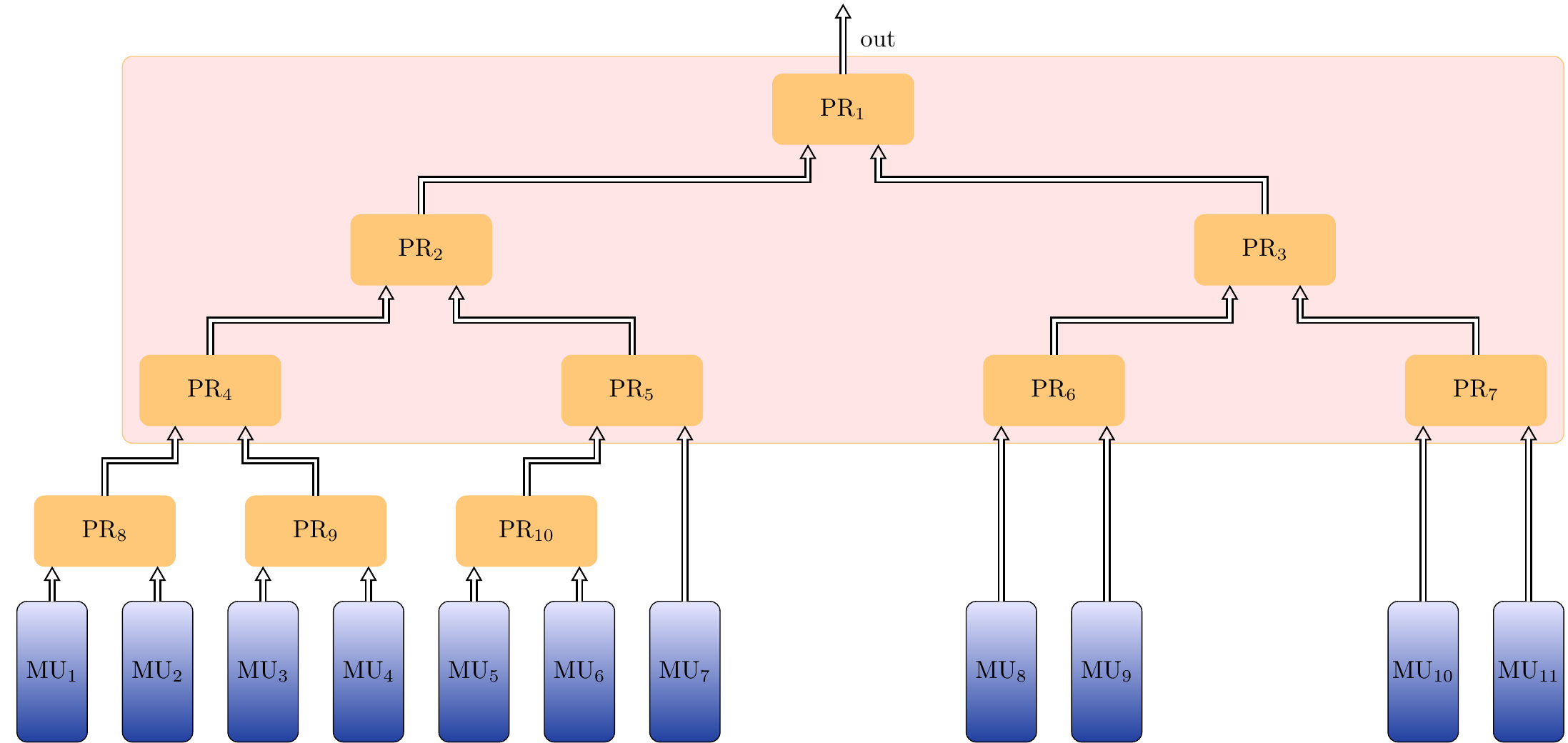}
    \caption{Schematic diagram of an input-extended incomplete binary-tree multiplexer (IIBTM). PR$_i$s and MU$_i$s denote binary photon routers and multiplexed units, respectively. Routers with a light red background form a 3-level complete binary-tree multiplexer (CBTM). Numbering of the PRs reflect the order in which they are added to the multiplexer.}
    \label{fig:branch_scheme}
\end{figure}

\begin{figure}
    \centering
    \includegraphics[width=\textwidth]{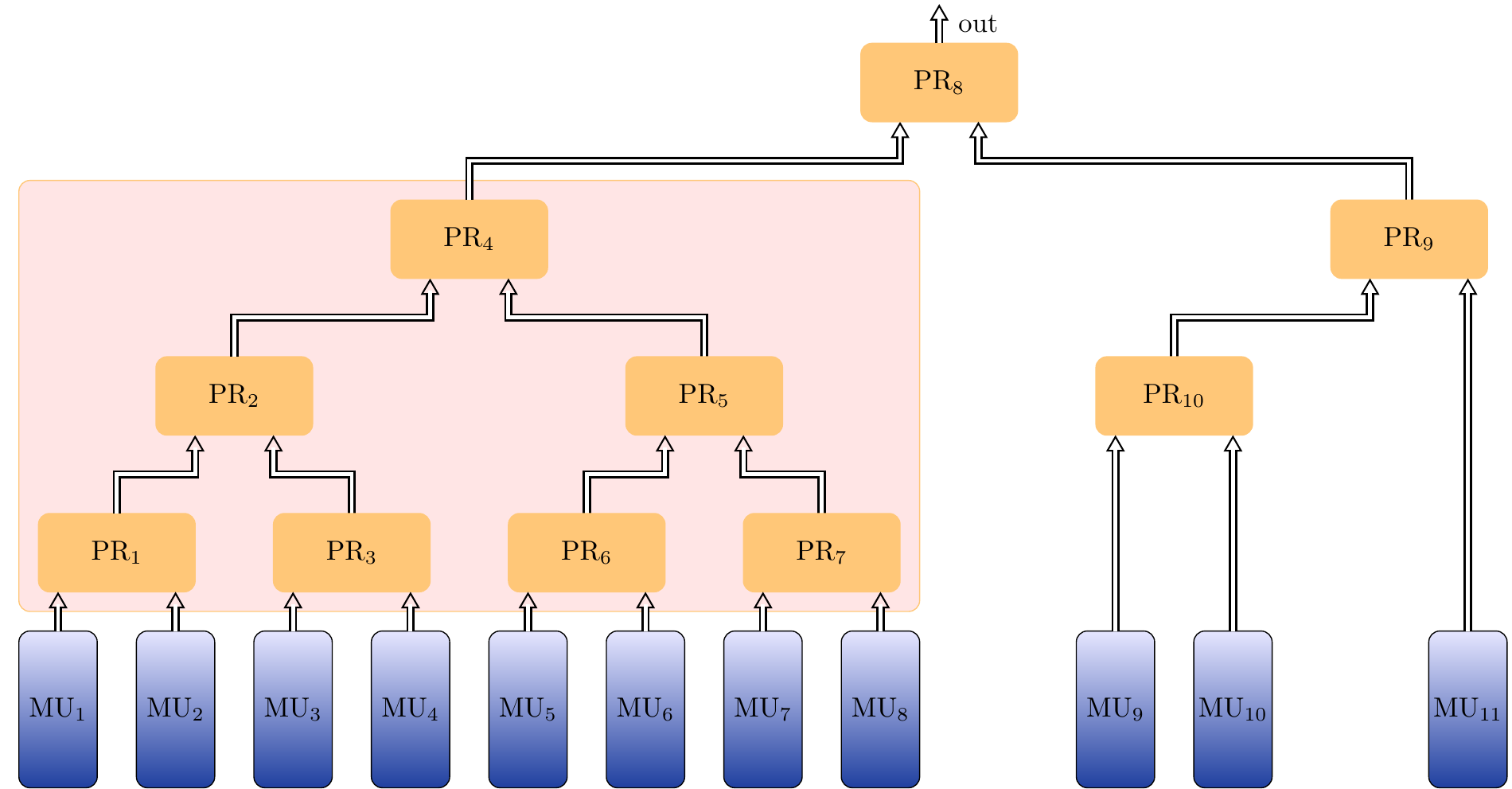}
    \caption{Schematic diagram of an output-extended incomplete binary-tree multiplexer (OIBTM). PR$_i$s and MU$_i$s denote binary photon routers and multiplexed units, respectively. Routers with a light red background form a 3-level complete binary-tree multiplexer (CBTM). Numbering of the PRs reflect the order in which they are added to the multiplexer.}
    \label{fig:root-scheme}
\end{figure}

An asymmetric architecture can have any number of inputs $N$.
However, in the case of the symmetric arrangement the number of inputs is restricted to a power of two, that is, $N=2^m$, where $m$ is the number of levels in the symmetric multiplexer.
Below we propose two novel arrangements presented in Figs.~\ref{fig:branch_scheme} and \ref{fig:root-scheme} that are essentially incomplete binary-tree multiplexers in which the number of inputs is arbitrary. In these schemes, an initially $m$-level symmetric multiplexer is extended step-by-step toward another, $m+1$-level symmetric multiplexer by adding new photon routers and multiplexed units to the system. In Figs.~\ref{fig:branch_scheme} and \ref{fig:root-scheme} PR$_i$s denote asymmetric binary (2-to-1) photon routers and MU$_i$s represent multiplexed units. The various inputs (or arms) of the multiplexers are numbered from left to right, their overall number $N$ is equal to the number of MUs.

The first proposed novel multiplexing scheme is presented in Fig.~\ref{fig:branch_scheme}. 
In this scheme, new asymmetric PRs are added to the inputs of an initially $m$-level symmetric multiplexer indicated by light red background in the figure one by one from left to right.
This building strategy is reflected by the numbering of the PRs in the figure.
This type of multiplexing scheme will be referred to as \emph{input-extended incomplete binary-tree multiplexer} (IIBTM).

The structure of the other proposed novel incomplete binary-tree multiplexer presented in Fig.~\ref{fig:root-scheme} is also based on an initially complete binary-tree multiplexer indicated by light red background in the figure.
The next step is to couple the output of the initial symmetric multiplexer into one of the inputs of a newly added router.
In the figure, such a novel router is indicated by PR$_8$.
Let us assume that always the left input of the novel router is used in such a situation.
Then another router is added to the other (right) input of the previously added router. In the figure, such a router is denoted by PR$_9$, and it is built onto the right input of PR$_8$.
Then the subsequent new routers are added to the incomplete branch of the multiplexer one by one from left to right until the given level is completed. This process is repeated until an $m+1$-level symmetric multiplexer is formed.
This building strategy is represented by the numbering of the PRs in the figure.
Throughout our paper, this arrangement will be referred to as \emph{output-extended incomplete binary-tree multiplexer} (OIBTM).

Next, we introduce the formulas characterizing the transmission through the various arms of the multiplexers. 
This quantity will be termed as \emph{total transmission coefficient} and denoted by $V_n$.
Its role is explained in detail in Sec.~\ref{sec:stat-theo}.

The formula describing the total transmission coefficient $V_n^{\rm sym}$ of the $n$th arm of a CBTM containing $m$ levels and $N=2^m$ inputs is
\begin{equation}
V_n^{\rm sym}=V_r^{m-H(n-1)}V_t^{H(n-1)}, \quad  n=[1,2,\dots,N],\label{eq:sym:seqnr}
\end{equation}
where $H(x)$ denotes the Hamming weight of $x$, that is, the number of ones in its binary representation.

In the case of an IIBTM let us assume that the overall number of inputs is $N$. In Fig.~\ref{fig:branch_scheme} this number is $N=11$. Denote the number of inputs or MUs at the level with the highest number by $N_1$. In the figure on level 4 there are 3 routers (PR$_8$ to PR$_{10}$) with 6 inputs, therefore $N_1=6$. Then the total transmission coefficient $V_n^{\rm in}$ characterizing the $n$th arm of an IIBTM can be expressed as
\begin{equation}
\begin{aligned}
V_n^{\rm in}&=V_r^{m_1+1-H(n-1)}V_t^{H(n-1)}&&\text{if}\quad 0<n\le N_1,\\
V_n^{\rm in}&=V_r^{m_1-H(n-N_1/2-1)}V_t^{H(n-N_1/2-1)}&& \text{if} \quad N_1<n\le N.
\end{aligned}\label{eq:branch:seqnr}
\end{equation}
Both quantities $m_1$ and $N_1$ are determined by the overall number of inputs $N$. The number of levels in the initial symmetric multiplexer is $m_1=\lfloor\log_2N\rfloor$, where $\lfloor x\rfloor$ denotes the floor function that gives as output the greatest integer less than or equal to $x$. In the figure $m_1=3$. Finally, the value $N_1$ can be derived as $N_1=2(N-2^{m_1})$.

As an example, we present the list of total transmission coefficients $V_n^{\rm in}$ characterizing the arms of the multiplexer of the IIBTM scheme shown in Fig.~\ref{fig:branch_scheme}:
\begin{equation}
V_n^{\rm in}=[V_r^4, V_r^3V_t, V_r^3V_t, V_r^2 V_t^2, V_r^3 V_t, V_r^2 V_t^2, V_r V_t^2, V_r^2 V_t, V_r V_t^2, V_r V_t^2, V_t^3].
\end{equation}

In the case of OIBTMs, assume again that the overall number of inputs of this multiplexer is $N$. In Fig.~\ref{fig:root-scheme} this number is $N=11$. Denote the number of inputs of the initial CBTM by $N_2$. In the figure, it is $N_2=8$. The number of inputs on the unfinished level under construction on the incomplete branch of the multiplexer below the output PR is denoted by $N_3$. In the figure $N_3=2$, that is, the number of inputs of PR$_{10}$. Then the total transmission coefficients $V_n^{\rm out}$ characterizing the arms of the OIBTM can be expressed as
\begin{equation}
\begin{array}{l@{\qquad}c@{\qquad}r@{\;}c@{\;}c@{\;}c@{\;}l}
V_n^{\rm out}=V_r^{m_2-H(n-1)}V_t^{H(n-1)},  &\text{if} & 0&<&n&\le& N_2\\
V_n^{\rm out}=V_tV_r^{m_3+1-H(n-N_2-1)}V_t^{H(n-N_2-1)},&\text{if} & N_2&<&n&\le& N_2+N_3\\
V_n^{\rm out}=V_tV_r^{m_3-H(n-N_2-\frac{N_3}{2}-1)}V_t^{H(n-N_2-\frac{N_3}{2}-1)}, &  \text{if} & N_2+N_3&<&n&\le& N.
\end{array}\label{eq:root:seqnr}
\end{equation}
All the quantities $m_i$ and $N_i$ can be derived from the overall number of inputs $N$. The value $m_2$ corresponding to the number of levels belonging to the branch of the OIBTM containing the initial symmetric multiplexer can be derived as $m_2=\lceil\log_2(N)\rceil$, where $\lceil x\rceil$ denotes the ceiling function that returns with the least integer greater than or equal to $x$. In the figure $m_2=4$.
Accordingly, $N_2$ can be expressed as $N_2=2^{m_2-1}$.
The number of finished levels, that is, the number of levels in the complete subtree on the incomplete branch of the multiplexer is $m_3=\lfloor\log_2(N-N_2)\rfloor$, where $\lfloor x\rfloor$ denotes the floor function.
In the figure, PR$_9$ itself forms a complete 1-level subtree, therefore $m_3=1$.
On level $m_3$, the number of inputs are $N_4=2^{m_3}$ ($N_4=2$ in the figure), but it may occur that new routers have been already added to some inputs on this level. Finally, the number of inputs on the next level of the incomplete branch is $N_3=2(N-N_2-N_4)$.

As an example, we show the total transmission coefficients $V_n^{\rm out}$ of the OIBTM presented in Fig.~\ref{fig:root-scheme}:
\begin{equation}
V_n^{\rm out}=[V_r^4,V_r^3V_t,V_r^3V_t,V_r^2V_t^2,V_r^3V_t,V_r^2V_t^2,V_r^2V_t^2,V_r V_t^3, V_r^2V_t,V_r V_t^2,V_t^2].
\end{equation}

Note that the total transmission coefficients presented in Eqs.~\eqref{eq:sym:seqnr}, \eqref{eq:branch:seqnr} and \eqref{eq:root:seqnr} are indexed according to their positions in the multiplexer, that is, the subsequent values are generally not sorted into an ascending or descending order.

\section{Statistical theory}\label{sec:stat-theo}

In order to analyze the proposed systems in detail we start from the general statistical theory introduced in Ref.~\cite{Bodog2020} that can be applied for describing any periodic SPSs based on spatial multiplexing and equipped with PNRDs. 
In this framework, it is assumed that $l$ photon pairs are generated in the $n$th multiplexed unit by a nonlinear source and the detection of a predefined number of photons $j$ ($j\leq l$) during a heralding event triggers the opening of the input ports of the multiplexer.
In general, $i$ signal photons can be expected at the output of the multiplexing system, the probability of which can be written as
\begin{eqnarray}
P_i^{(S)}=\big(1-\sum_{j\in S}P^{(D)}(j)\big)^N\delta_{i,0}+\sum_{n=1}^N\left[\big(1-\sum_{j\in S}P^{(D)}(j)\big)^{n-1}\times\sum_{l=i}^\infty\sum_{j\in S} P^{(D)}(j|l)P^{(\lambda)}(l)V_n(i|l)\right].\label{general_formula}
\end{eqnarray}
In this formula $P_n^{(D)}(j)$ denotes the probability of detecting exactly $j$ photons in the $n$th multiplexed unit.
$P^{(D)}(j|l)$ is the conditional probability of registering $j$ photons, provided that $l$ photons arrive at the detector.
The probability of generating $l$ photon pairs in the $n$th multiplexed unit when the mean photon number of the generated photon pairs is $\lambda_n$ in that unit is denoted by  $P_n^{(\lambda_n)}(l)$.
$V_n(i|l)$ stands for the conditional probability of the event that the output of the multiplexer is reached by $i$ photons, provided that the number of signal photons arriving from the $n$th multiplexed unit into the system is $l$.
The set $S$ describes the detection strategy, that is, it contains elements from the set of positive integers $\mathbb{Z}^+$ up to a predefined boundary value $J_b$.

The first term in Eq.~\eqref{general_formula} contributes only to the case where no photon reaches the output, that is, to the probability $P_0^{(S)}$.
It describes the case when none of the PNRDs in the multiplexed units have detected a photon number in $S$.
The second term in Eq.~\eqref{general_formula} describes the case that the heralding event occurs in the $n$th unit.
The first factor in this term is the probability that none of the first $n-1$ detectors have detected a photon number in $S$.
The second factor corresponds to the event that out of the $l$ photons entering the multiplexer from the $n$th multiplexed unit after heralding, only $i$ reach the output due to the losses of the multiplexer. The summation over $n$ in the second term takes into consideration all the possible contributions to the probability $P_i^{(S)}$.

In Eq.~\eqref{general_formula} the various probabilities can be expressed as follows. The probability $P^{(\lambda)}(l)$ represents that a nonlinear source generates $l$ photon pairs. In our calculations we use Poisson distribution, that is,
\begin{equation}
P^{(\lambda)}(l)=\frac{\lambda^l e^{-\lambda}}{l!},
\end{equation}
$\lambda$ representing the mean photon number of the photon pairs generated by the nonlinear source and arriving at the detectors in the multiplexed units, that is, the input of the heralding process. Hence, we refer to it by the term \emph{input mean photon number} in the following. Poissonian distribution is valid for multimode SPDC or SFWM processes, that is, for weaker spectral filtering \cite{Avenhaus2008, Mauerer2009, Takesue2010, Almeida2012, Collins2013, Harder2016, KiyoharaOE2016}. Assuming this distribution makes it possible to compare our results with the ones presented in a significant part of the literature related to SPS, which were also obtained for Poissonian distribution~\cite{Jeffrey2004, ShapiroWong2007, Ma2010, Mower2011, MazzarellaPRA2013, Adam2014, Bodog2016, MagnoniQIP2019}.
The formula for the conditional probability $P^{(D)}(j|l)$ can be obtained as
\begin{eqnarray}
P^{(D)}(j|l)=\binom{l}{j}V_D^j(1-V_D)^{l-j},
\end{eqnarray}
where $l$ and $j$ are the number of the generated and detected photons inside a multiplexed unit, respectively, by using a detector with efficiency $V_D$.
Finally, the total probability $P^{(D)}(j)$ reads
\begin{eqnarray}
P^{(D)}(j)=\sum_{l=j}^\infty P^{(D)}(j|l)P^{(\lambda)}(l).
\end{eqnarray}
The conditional probability 
\begin{equation}
V_n(i|l)=\binom{l}{i}V_n^i(1-V_n)^{l-i}\label{eq:Vn:cond}
\end{equation}
describes the case when $i$ signal photons reach the output of the whole multiplexer given that $l$ signal photons are generated in the $n$th multiplexed unit.
For analyzing a particular setup the corresponding total transmission coefficient $V_n$ should be substituted into Eq.~\eqref{eq:Vn:cond}.
In the case of the schemes studied in this paper the formulas in Eqs.~\eqref{eq:sym:seqnr}, \eqref{eq:branch:seqnr}, and \eqref{eq:root:seqnr} can be used for the CBTM, the IIBTM, and the OIBTM schemes, respectively.
We note that 
for the priority logic that corresponds to Eq.~\eqref{general_formula} the preferred multiplexed unit is the one with the smallest sequential number $n$ when heralding events occur in multiple units.
Therefore, it is appropriate to reorder the numbering of the multiplexed units in these equations so that the corresponding total transmission coefficients $V_n$ are arranged into a decreasing order.
The numbering of the multiplexer arms having identical total transmission coefficients $V_n$ is arbitrary.
Applying such an indexing the multiplexer arm with the highest $V_n$ corresponding to the smallest loss is preferred by the priority logic when multiple heralding events occur that can result in higher single-photon probabilities.

Using the presented framework, the optimization of the considered systems can be accomplished in the following way.
We fix the total transmission coefficients $V_n$ of the systems, that is, $V_n^{\rm sym}$, $V_n^{\rm in}$, and $V_n^{\rm out}$, for the CBTM, the IIBTM, and the OIBTM, respectively. This means that the reflection and transmission efficiencies of the router $V_r$ and $V_t$, respectively, and the general transmission coefficient $V_b$ are fixed. We also fix the detection strategy $S$ and the detector efficiency $V_D$.
Then two parameters are left to be optimized that are the input mean photon number $\lambda$ and the number of multiplexed units $N$.
First, we determine the optimal value of the input mean photon number $\lambda$ for which the single-photon probability $P_1$ is the highest for a fixed value of the number of multiplexed units $N$.
This probability is termed as the \emph{achievable single-photon probability} and denoted by $P_{1,N}$ while the corresponding photon number is called \emph{optimal input mean photon number} and denoted by $\lambda_\opt$.
We repeat this procedure for all reasonable values of the number of multiplexed units $N$ and select the optimal value $N_{\opt}$ for which the achievable single-photon probability $P_{1,N}$ is the highest. 
This probability is termed as the \emph{maximal single-photon probability}, denoted by $P_{1,\max}$ and it belongs to the \emph{optimal number of multiplexed units} $N_\opt$ and the optimal input mean photon number $\lambda_\opt$ corresponding to $N_\opt$.
We remark that, although $\lambda_\opt$ can be found for every value of $N$, we do not use a separate term for the one belonging to the optimal number of multiplexed units $N_\opt$.

We note that in the following we use the superscripts ``out'', ``in'', and ``sym'' for all these quantities to denote results achieved for SPSs based on OIBTM, IIBTM, and CBTM, respectively.
\section{Results}
In this section, we summarize our results regarding the optimization of SPSs based on the proposed incomplete binary-tree multiplexers composed of asymmetric routers.
Asymmetric routers with high transmission efficiencies can be realized with bulk optical elements, as it is presented in Fig.~\ref{fig:Router}.
Therefore, the ranges of the various transmission efficiencies are chosen so that their upper boundaries correspond to the best loss parameters of state-of-the-art bulk optical elements.
Accordingly, the upper boundaries of the ranges of the reflection and transmission efficiencies of the router are taken to be $V_r=0.99$ and $V_t=0.985$, respectively \cite{Peters2006, Kaneda2019}. For the detector efficiency, we choose $V_D=0.98$ as the upper boundary because this value is the highest one reported in the literature \cite{Fukuda2011}.
The general transmission coefficient $V_b$ is strongly affected by the actual experimental realization of the system; we assume the value of $V_b=0.98$ for its highest feasible value.

\begin{figure}[!b]
\begin{center}
\includegraphics[width=0.48\textwidth]{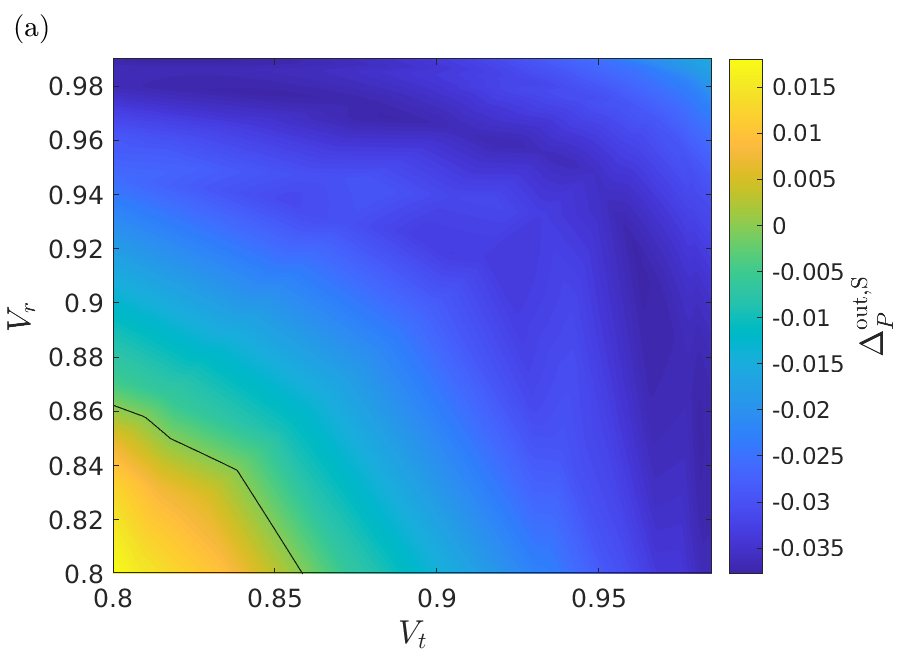}
\includegraphics[width=0.48\textwidth]{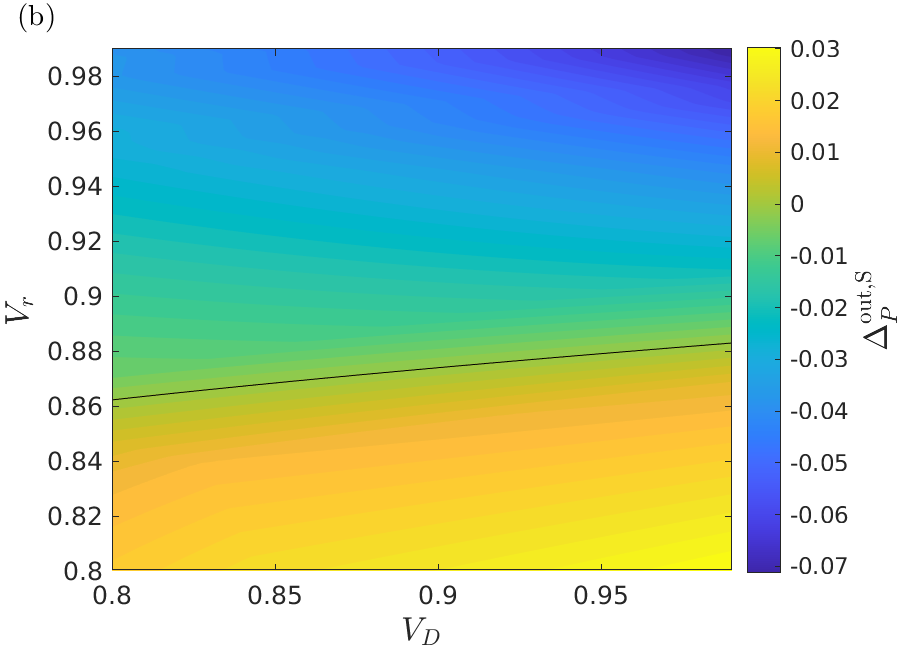}
\end{center}
\caption{(a) The difference $\Delta_P^{\text{out}, S} = P_{1,\max}^{\text{out},S=\{1,2\}}- P_{1,\max}^{\text{out},\SPD}$ between the maximal single-photon probabilities for SPSs based on OIBTM obtained by assuming $S=\{1,2\}$ detection strategy and SPD strategy, respectively, as a function of the transmission efficiency $V_t$ and the reflection efficiency $V_r$ for the detector efficiency $V_D=0.8$ and the general transmission coefficient $V_b=0.9$.
(b) The same quantity $\Delta_P^{\text{out}, S}$ as a function of the detector efficiency $V_D$ and the reflection efficiency $V_r$ for the transmission efficiency $V_t=0.8$ and the general transmission coefficient $V_b=0.9$. Below the continuous lines, the detection strategy $S=\{1,2\}$ outperforms the SPD strategy.\label{fig:3}}
\end{figure}

Let us first clarify the role of the detection strategy.
First, we determine the ranges of the loss parameters for which the maximal single-photon probability $P_{1,\max}$ achieved with SPSs based on the proposed multiplexing schemes and applying the $S=\{1,2\}$ detection strategy surpasses the same probability achieved by applying the SPD strategy.
Fig.~\ref{fig:3}(a) shows the difference $\Delta_P^{\text{out}, S} = P_{1,\max}^{\text{out},S=\{1,2\}}- P_{1,\max}^{\text{out},\SPD}$ between the maximal single-photon probabilities for SPSs based on OIBTM obtained by assuming $S=\{1,2\}$ detection strategy and SPD strategy, respectively, as a function of the transmission efficiency $V_t$ and the reflection efficiency $V_r$ for the detector efficiency $V_D=0.8$ and the general transmission coefficient $V_b=0.9$.
Below the continuous line, that is, for smaller values of the reflection and transmission efficiencies $V_r$ and $V_t$, respectively, the detection strategy $S=\{1,2\}$ outperforms the SPD strategy.

Fig.~\ref{fig:3}(b) shows the same quantity $\Delta_P^{\text{out}, S}$ as a function of the detector efficiency $V_D$ and the reflection efficiency $V_r$ for the transmission efficiency $V_t=0.8$ and the general transmission coefficient $V_b=0.9$.
Below the line, the detection strategy $S=\{1,2\}$ outperforms the SPD strategy.
The figure shows that this difference does not have a strong dependence on the detector efficiency $V_D$.
We note that increasing the general transmission coefficient $V_b$ would cause the level $\Delta_P^{\text{out}, S}=0$ to shift toward smaller values of the reflection and transmission efficiencies $V_r$ and $V_t$, respectively.
We calculated similar differences $\Delta_P^{\text{in},S}$ for SPSs based on IIBTM and $\Delta_P^{\text{sym},S}$ for SPSs based on CBTM. In these cases, we also found that the $S=\{1,2\}$ strategy outperforms the SPD strategy only for smaller values of the reflection and transmission efficiencies $V_r$ and $V_t$, respectively. Therefore, in our analysis we use the lower boundary 0.9 of the ranges for the efficiencies $V_r$, $V_t$, and $V_D$ ensuring that SPD is the optimal detection strategy for the whole considered parameter ranges. As all the subsequent results are obtained for the SPD strategy, henceforth we do not indicate this fact.

\begin{figure}
    \centering
    \includegraphics[width=0.48\textwidth]{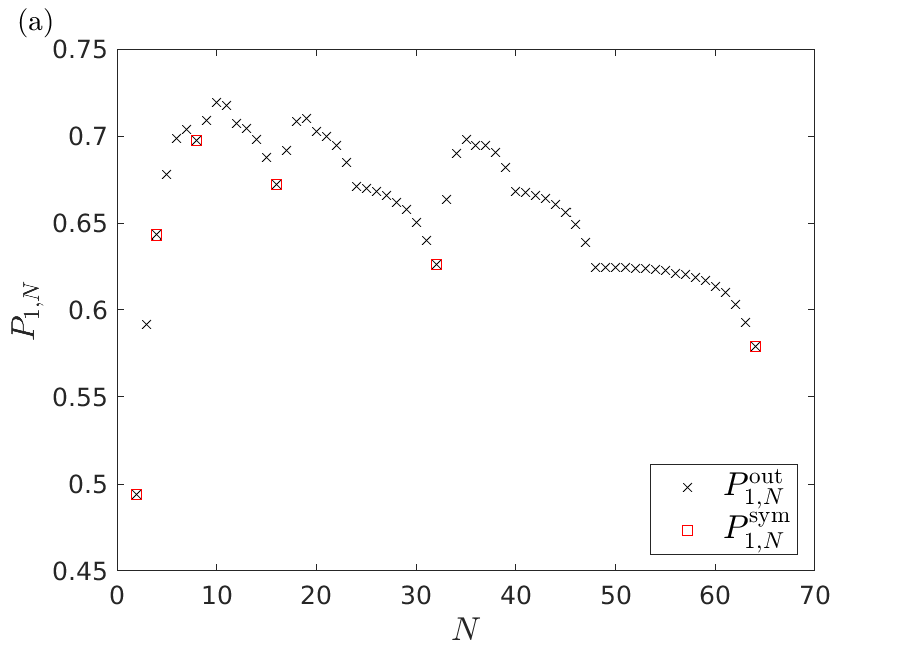}
    \includegraphics[width=0.48\textwidth]{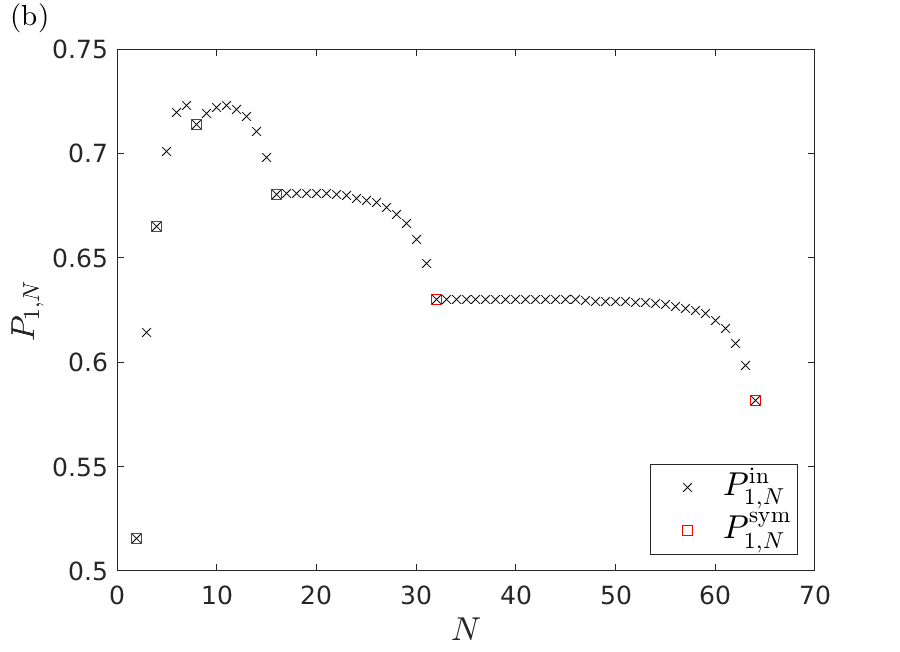}
    \caption{The achievable single-photon probabilities $P_{1,N}$ as a function of the number of multiplexed units $N$ for SPSs based on (a) OIBTM for the parameters $V_b=0.98$, $V_D=0.9$, $V_r=0.92$, and $V_t=0.9$, and on (b) IIBTM for the parameters $V_b=0.98$, $V_D=0.95$, $V_r=0.92$, and $V_t=0.9$. For comparison, the same quantity is presented for CBTM, denoted by red squares.}
    \label{fig:root_Ncompare}
\end{figure}
Next, we discuss the properties of the achievable single-photon probabilities for the proposed schemes.
Figure~\ref{fig:root_Ncompare} shows typical results for the achievable single-photon probabilities $P_{1,N}$ as a function of the number of multiplexed units $N$. In Fig.~\ref{fig:root_Ncompare}(a) we present results for SPSs based on OIBTM for the general transmission coefficient $V_b=0.98$, the detector efficiency $V_D=0.9$, the reflection efficiency $V_r=0.92$, and the transmission efficiency $V_t=0.9$, while Figure~\ref{fig:root_Ncompare}(b) is an example for SPSs based on IIBTM for the parameters $V_b=0.98$, $V_D=0.95$, $V_r=0.92$, and $V_t=0.9$.
Note that CBTMs are special cases of incomplete binary-tree multiplexers for certain values of the number of multiplexed units, therefore the point sequences presented in Fig.~\ref{fig:root_Ncompare} contain results for CBTM also.
These points are marked with red squares.
In previous studies \cite{Adam2014, Bodog2020} it was found that for CBTM, the achievable single-photon probability $P_{1,N}$ as a function of the number of multiplexed units $N$ has a single maximum.
This is due to the fact that increasing the system size, that is, the number of levels in the multiplexer the losses in the system are increased, that is, all the total transmission coefficients $V_n$ assigned to the various branches of the multiplexer are decreased that deteriorates the benefit of multiplexing.
However, the achievable single-photon probabilities presented in Fig.~\ref{fig:root_Ncompare} for SPSs based on OIBTM and IIBTM show local maxima for values of the number of multiplexed units $N$ that are between the special power-of-two numbers of multiplexed units characterizing the CBTM.
The absolute maxima of the achievable single-photon probabilities presented in Fig.~\ref{fig:root_Ncompare} for SPSs based on incomplete binary-tree multiplexers are higher than the maximum of the same quantity for SPSs based on CBTM.
These maxima can occur either for smaller or larger values of the number of multiplexed units $N$ than for CBTM.
As no simple rule can be found for this behavior with respect to the parameters $V_b$, $V_D$, $V_r$, and $V_t$, we did not analyze this problem in detail.
We note that the breaking points in the point sequences representing OIBTM in Fig.~\ref{fig:root_Ncompare}(a) correspond to the case when a new router is added to a complete symmetric subtree on the incomplete branch of the OIBTM.
However, in the case of IIBTM in Fig.~\ref{fig:root_Ncompare}(b) the breaking points can be observed only when a new router is added to a CBTM.

\begin{figure}[!tb]
\centering
\includegraphics[width=0.48\textwidth]{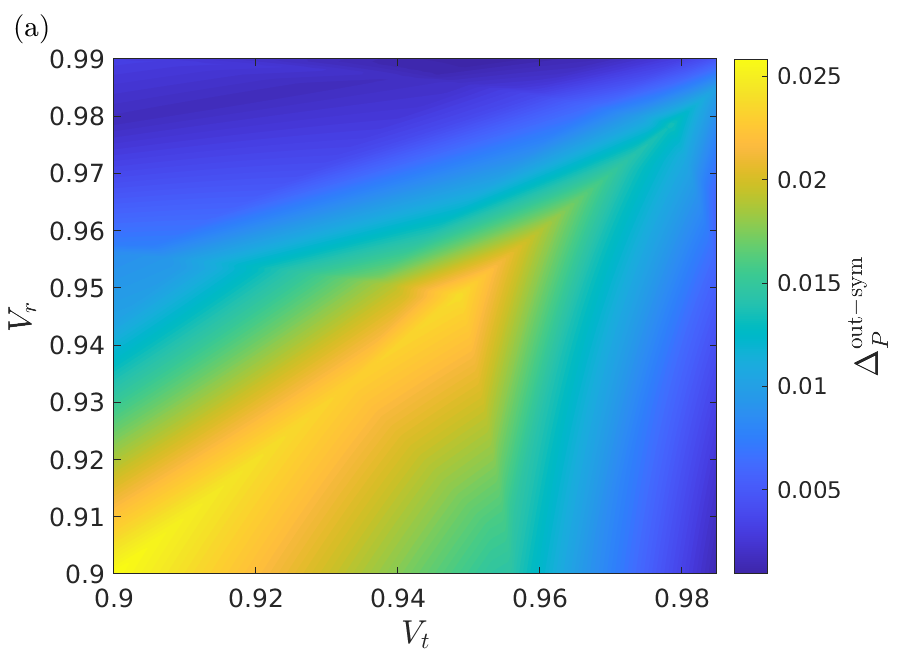}
\includegraphics[width=0.48\textwidth]{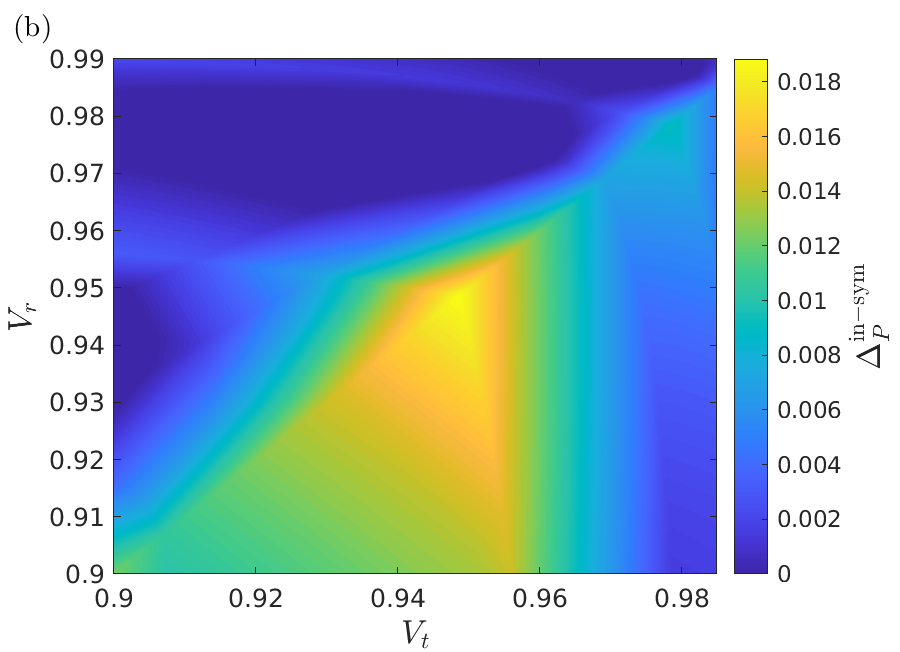}
\caption{(a) The difference $\Delta_P^{\text{out}-\text{sym}}=P_{1,\max}^{\text{out}}-P_{1,\max}^{\text{sym}}$ between the maximal single-photon probabilities for SPSs based on OIBTM and on CBTM, respectively, as a function of the transmission efficiency $V_t$ and the reflection efficiency $V_r$ for the general transmission coefficient $V_b=0.98$ and the detector efficiency $V_D=0.9$.
(b) The corresponding function $\Delta_P^{\text{in}-\text{sym}}=P_{1,\max}^{\text{in}}-P_{1,\max}^{\text{sym}}$ for SPSs based on IIBTM and on CBTM, respectively.}
\label{fig:P1diffs}
\end{figure}

Next, we compare the performance of the two proposed incomplete binary-tree multiplexer schemes and that of the CBTM scheme. The results of these calculations are presented in Fig.~\ref{fig:P1diffs}.
Figure~\ref{fig:P1diffs}(a) shows the difference $\Delta_P^{\text{out}-\text{sym}}=P_{1,\max}^{\text{out}}-P_{1,\max}^{\text{sym}}$ between the maximal single-photon probabilities for SPSs based on OIBTM and on CBTM, respectively, as a function of the transmission efficiency $V_t$ and the reflection efficiency $V_r$ for the general transmission coefficient $V_b=0.98$ and the detector efficiency $V_D=0.9$.
The corresponding function  $\Delta_P^{\text{in}-\text{sym}}=P_{1,\max}^{\text{in}}-P_{1,\max}^{\text{sym}}$ for SPSs based on IIBTM and on CBTM, respectively, can be seen in Fig.~\ref{fig:P1diffs}(b).
The figures show that SPSs based on either OIBTM or IIBTM outperform SPSs based on CBTM on the considered range of the parameters.
It is in accordance with the expectations, as CBTMs are special cases of OIBTMs and IIBTMs.
The advantage introduced by the OIBTM or the IIBTM is smaller for higher values of the reflection efficiencies $V_r$ or the transmission efficiencies $V_t$.
From the figures one can also deduce that for a fixed value of the transmission efficiency $V_t$ or the reflection efficiency $V_r$ the highest differences $\Delta_P^{\text{out}-\text{sym}}$ or $\Delta_P^{\text{in}-\text{sym}}$ can be obtained for SPSs based on multiplexers equipped with symmetric routers, that is, when $V_r=V_t$.
The maximal value of the difference $\Delta_{P,\max}^{\text{out}-\text{sym}}$ is $0.026$ that can be achieved at $V_r=V_t=0.9$ while the maximal value of the difference $\Delta_P^{\text{in}-\text{sym}}$ is $0.019$ that occurs for the values $V_r=V_t=0.949$.
Note that the details of the functions of Figs.~\ref{fig:P1diffs} including the particular data of these maxima are affected by the actual values of the detector efficiency $V_D$ and the general transmission coefficient $V_b$. However, the main characteristics of these functions remain the same.
We also remark that, due to the asymmetry of both OIBTMs and IIBTMs, the images presented in Fig.~\ref{fig:P1diffs} are asymmetric.

\begin{figure}[!tb]
\centering
\includegraphics[width=0.48\textwidth]{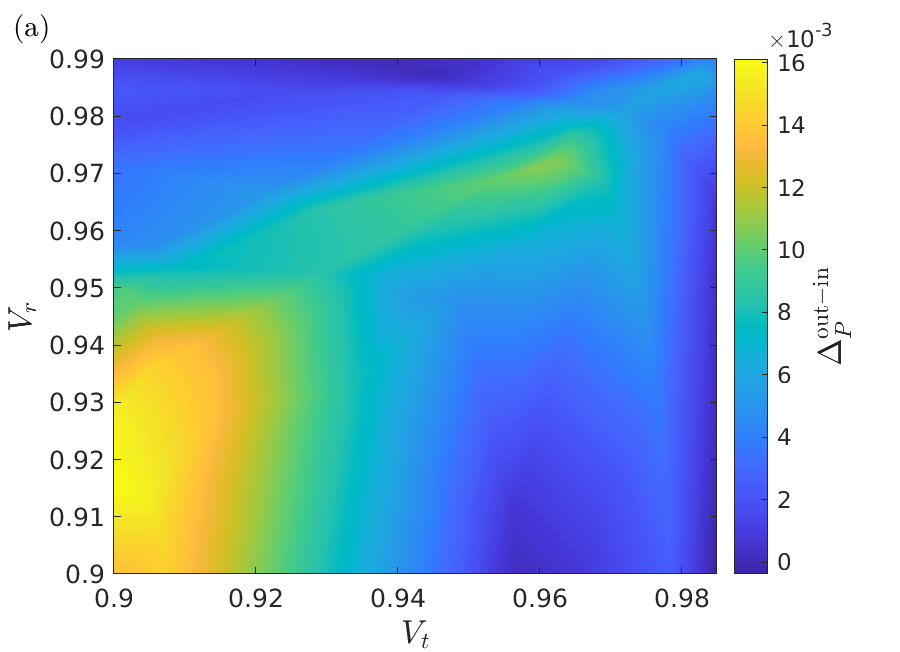}
\includegraphics[width=0.48\textwidth]{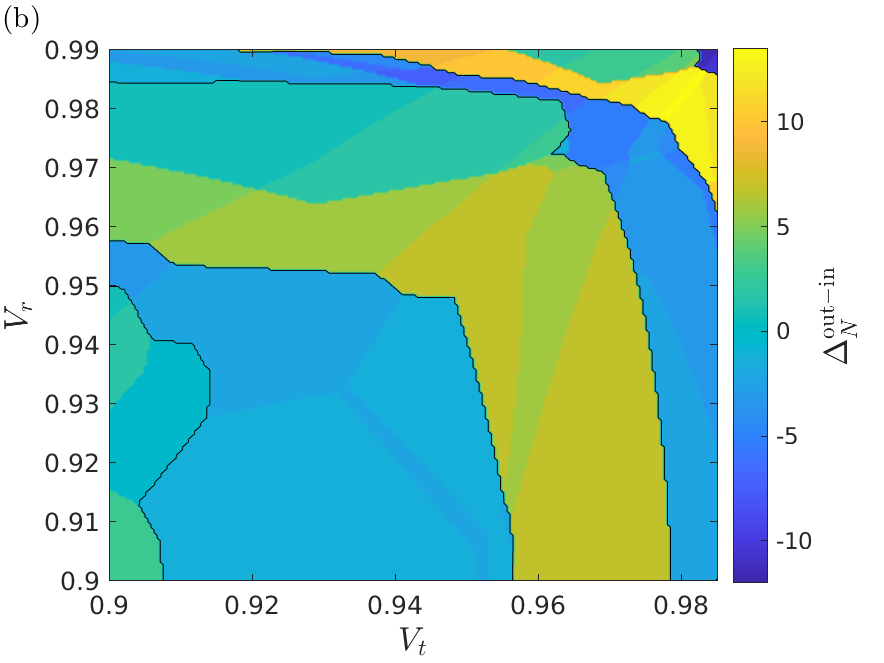}
\caption{
(a) The difference $\Delta_P^{\text{out}-\text{in}}=P_{1,\max}^{\text{out}}-P_{1,\max}^{\text{in}}$ between the maximal single-photon probabilities for SPSs based on OIBTM and on IIBTM, respectively, as a function of the transmission efficiency $V_t$ and the reflection efficiency $V_r$ for the general transmission coefficient $V_b=0.98$ and the detector efficiency $V_D=0.9$.
(b) The corresponding function for the difference $\Delta_N^{\rm out-in}=N_{\opt}^{\rm out}-N_{\opt}^{\rm in}$ between the optimal number of multiplexed units.
Continuous black lines separate regions where OIBTM or IIBTM outperforms the other with respect to the optimal number of multiplexed units.}
\label{fig:out-in}
\end{figure}

Finally, we compare the performance of the two novel proposed incomplete binary-tree multiplexer schemes.
In Fig.~\ref{fig:out-in}(a) we show the difference $\Delta_P^{\text{out}-\text{in}}=P_{1,\max}^{\text{out}}-P_{1,\max}^{\text{in}}$ between the maximal single-photon probabilities for SPSs based on OIBTM and on IIBTM, respectively, as a function of the transmission efficiency $V_t$ and  the reflection efficiency $V_r$ for the general transmission coefficient $V_b=0.98$ and the detector efficiency $V_D=0.9$.
According to the figure, OIBTM outperforms IIBTM in the whole considered ranges of the transmission and reflection efficiencies $V_t$ and $V_r$, respectively.
Intuitively, this observation can be explained as follows.
In the case of IIBTM, the addition of new routers to the multiplexer always leads to new branches with higher losses, that is, smaller total transmission efficiencies $V_n$ than the ones in the initial CBTM.
On the contrary, for OIBTM, the total transmission efficiencies $V_n$ of the novel branches of OIBTM are always higher than the ones characterizing the initial CBTM.
As we described in Sec.~\ref{sec:stat-theo}, the priority logic prefers multiplexer arms with higher $V_n$ in the case of multiple heralding events, therefore this property of OIBTM can result in higher single-photon probabilities.
For the values of the general transmission coefficient $V_b$ and the detector efficiency $V_D$ used in these calculations, the highest differences between the maximal single-photon probabilities $P_{1,\max}$ of the two schemes are $\Delta_P^{\text{out}-\text{in}}\approx 0.016$ at $V_t\approx 0.9$ and  $V_r\approx0.92$.

The results thus far showed that by using incomplete binary-tree multiplexers, it is possible to increase the maximal single-photon probability.
However, from an experimental point of view, the number of optical elements required to realize these multiplexers is also important.
In Fig.~\ref{fig:out-in}(b) we show the difference $\Delta_N^{\rm out-in}=N_{\opt}^{\rm out}-N_{\opt}^{\rm in}$ between the optimal number of multiplexed units for SPSs based on OIBTM and on IIBTM, respectively, as a function of the transmission efficiency $V_t$ and the reflection efficiency $V_r$ for the general transmission coefficient $V_b=0.98$ and the detector efficiency $V_D=0.9$.
The figure shows that the difference $\Delta_N^{\rm out-in}$, that is, the experimentally optimal choice of the multiplexing scheme, is strongly affected by the efficiencies $V_t$ and  $V_r$ of the routers.
The difference fluctuates between positive and negative values, that is, for some parameters the number of multiplexed units is smaller for SPSs based on OIBTM, for other parameters this quantity is smaller for SPSs based on IIBTM.
For small values of the reflection and transmission efficiencies $V_r$ and $V_t$, respectively, the absolute difference $\left|\Delta_N^{\rm out-in}\right|$ is rather small while for high values of these efficiencies $\left|\Delta_N^{\rm out-in}\right|$ increases.
The difference between the optimal number of multiplexed units can be as high as $\Delta_N^{\rm out-in}\approx15$.
Note, however, that in these cases the optimal number of multiplexed units $N_{\opt}$ is also very high ($N_{\opt}\approx 40$).
In view of these observations, when an experiment is realized with finite experimental resources with given loss parameters, it is worth using the full statistical treatment presented in this paper to determine which of the proposed incomplete multiplexers yields higher maximal single-photon probability with less optical elements.

\begin{table*}[!tb]
\caption{Maximal single-photon probabilities $P^{\text{out}}_{1,\max}$ for SPS based on OIBTM, the required optimal number of multiplexed units $N_\opt^{\text{out}}$, and the optimal input mean photon numbers $\lambda_\opt^{\text{out}}$ at which they can be achieved for various values of the reflection efficiency $V_r$, the transmission efficiency $V_t$ and the detector efficiency $V_D$, and for the general transmission coefficient $V_b=0.98$. \label{tab:2}}
\centerline{
\begin{tabular}{c|c|ccc|ccc|ccc}\hline
      &&\multicolumn{3}{c|}{$V_t=0.9$}
      & \multicolumn{3}{c|}{$V_t=0.95$}
      & \multicolumn{3}{c}{$V_t=0.985$}\\\hline
$V_r$& $V_D$ &
$P_{1,\max}^{\text{out}}$&$N_\opt^{\text{out}}$&$\lambda_{\opt}^{\text{out}}$&
$P_{1,\max}^{\text{out}}$&$N_\opt^{\text{out}}$&$\lambda_{\opt}^{\text{out}}$&
$P_{1,\max}^{\text{out}}$&$N_\opt^{\text{out}}$&$\lambda_{\opt}^{\text{out}}$
\\\hline
0.92 & 0.8  & 0.685 & 10 & 0.686 & 0.743 & 20 & 0.446 & 0.809 & 40 & 0.315 \\
0.92 & 0.9  & 0.716 & 10 & 0.78  & 0.772 & 11 & 0.696 & 0.835 & 20 & 0.517 \\
0.92 & 0.95 & 0.733 & 10 & 0.869 & 0.793 & 10 & 0.836 & 0.855 & 20 & 0.658 \\
0.92 & 0.98 & 0.744 & 10 & 0.943 & 0.808 & 10 & 0.925 & 0.87  & 20 & 0.824 \\\hline
0.97 & 0.8  & 0.757 & 17 & 0.472 & 0.801 & 36 & 0.262 & 0.862 & 40 & 0.205 \\
0.97 & 0.9  & 0.787 & 17 & 0.576 & 0.828 & 18 & 0.466 & 0.88  & 38 & 0.279 \\
0.97 & 0.95 & 0.805 & 17 & 0.711 & 0.845 & 18 & 0.586 & 0.896 & 20 & 0.464 \\
0.97 & 0.98 & 0.818 & 9  & 0.927 & 0.858 & 10 & 0.87  & 0.908 & 19 & 0.682 \\\hline
0.99 & 0.8  & 0.807 & 34 & 0.324 & 0.852 & 40 & 0.214 & 0.899 & 74 & 0.114 \\
0.99 & 0.9  & 0.834 & 18 & 0.513 & 0.872 & 33 & 0.314 & 0.911 & 37 & 0.213 \\
0.99 & 0.95 & 0.854 & 17 & 0.66  & 0.888 & 17 & 0.534 & 0.921 & 36 & 0.269 \\
0.99 & 0.98 & 0.869 & 17 & 0.82  & 0.901 & 17 & 0.692 & 0.931 & 18 & 0.561\\\hline
\end{tabular}}
\end{table*}

Let us now assess the performance of SPSs based on OIBTM in more detail.
Table \ref{tab:2} shows the maximal single-photon probabilities $P^{\text{out}}_{1,\max}$ for this setup, the required optimal number of multiplexed units $N_\opt^{\text{out}}$, and the optimal input mean photon numbers $\lambda_\opt^{\text{out}}$ at which they can be achieved for various values of the reflection efficiency $V_r$, the transmission efficiency $V_t$ and the detector efficiency $V_D$, and for the general transmission coefficient $V_b=0.98$.

From the table one can deduce that increasing any of the three parameters $V_r$, $V_t$, or $V_D$ leads to an increase in the single-photon probability $P^{\text{out}}_{1,\max}$.
The highest single-photon probabilities that can in principle be achieved by output-extended systems is higher than 0.93 for the parameters $V_b=0.98$, $V_D=0.98$, $V_r=0.99$ and $V_t=0.985$. These parameters are considered to be realizable by state-of-the-art technology.
At a given value of $V_D$ the increase of the reflection and transmission efficiencies $V_r$ and $V_t$, respectively, is generally accompanied by an increase in the required number of multiplexed units $N_\opt^{\text{out}}$.
Obviously, smaller losses corresponding to higher transmissions allow us to use larger optimal system sizes to achieve higher maximal single-photon probabilities $P^{\text{out}}_{1,\max}$ via multiplexing. 
However, an increase in the detector efficiency $V_D$ corresponding to an increased probability that the single-photon events are selected correctly by the detectors generally leads to a decrease in $N_\opt^{\text{out}}$. This is due to the fact that in this case the multi-photon events can be excluded by the detectors themselves, there is no need for suppressing the occurrence of these events by decreasing the intensity and subsequently increasing the system size, that is, introducing longer arms with higher losses to the multiplexer is not so crucial anymore.

The observations concerning the optimal input mean photon number $\lambda_{\opt}^{\rm out}$ are the opposite. Increasing either the reflection efficiency $V_r$ or the transmission efficiency $V_t$ the optimal input mean photon number $\lambda_{\opt}^{\rm out}$ decreases, while increasing the values of the detector efficiency $V_D$ leads to an increase in the values of $\lambda_{\opt}^{\rm out}$. The finding that the relationship between the optimal number of multiplexed units $N_\opt^{\text{out}}$ and the optimal input mean photon number $\lambda_{\opt}^{\rm out}$ is inverse is not unexpected: having less multiplexed units in the multiplexing system can be compensated by higher values of the input mean photon numbers that guarantees the occurrence of at least one heralding event in the whole multiplexer.

\section{Conclusion}
We have proposed two types of incomplete binary-tree multiplexers aiming at increasing the performance of spatially multiplexed single-photon sources.
These multiplexers contain incomplete branches either at the input or at the output of the symmetric ones, hence the power-of-two restriction on the number of multiplexed units characterizing symmetric multiplexers is eliminated for them.
We applied a general statistical theory that includes all relevant loss mechanisms for analyzing and optimizing these single-photon sources based on these multiplexers realized with general asymmetric routers and photon-number-resolving detectors.
We have shown that the use of any of the two proposed multiplexing systems can lead to higher single-photon probabilities than that achieved with complete binary-tree multiplexers applied thus far in experiments.
We have found that the performance of single-photon sources based on output-extended incomplete binary-tree multiplexers is better than that of those based on input-extended ones in the considered ranges of the parameters.
The single-photon probabilities that can in principle be achieved by output-extended systems are higher than 0.93 when they are realized by state-of-the-art bulk optical elements.

\begin{backmatter}
\bmsection{Funding}
National Research, Development and Innovation Office, Hungary (Project No.\ K124351 and the Quantum Information National Laboratory of Hungary); European Union (Grants No.\ EFOP-3.6.1.-16-2016-00004, No.\ EFOP-3.6.2-16-2017-00005 and No.\ EFOP-3.4.3-16-2016-00005).
\end{backmatter}

\end{document}